\documentstyle[12pt,aasms4]{article}

\def\beq{\begin{equation}}
\def\eeq{\end{equation}}
\def\bey{\begin{eqnarray}}
\def\eey{\end{eqnarray}}
\def\kms{\mbox{\rm \,km\,s}^{-1}}
\def\kpc{\;\rm kpc}
\def\msun{\;{\rm M}_\odot}
\def\solmasspc{\;{\msun {\rm pc}{}^{-2}}}
\def\MLMC{M_{\rm LMC}}
\def\Mb{M_{\rm b}}
\def\Md{M_{\rm d}}
\def\NLMC{N_0}
\def\fb{f_{\rm b}}
\def\Id{I_{\rm d}}
\def\Ib{I_{\rm b}}
\def\Ld{L_{\rm d}}
\def\Lb{L_{\rm b}}
\def\Wd{W_{\rm d}}
\def\Wb{W_{\rm b}}
\def\hd{\Delta_{\rm d}}
\def\hb{\Delta_{\rm b}}
\def\xd{X_{\rm d}}
\def\yd{Y_{\rm d}}
\def\xb{X_{\rm b}}
\def\yb{Y_{\rm b}}
\def\spose#1{\hbox to 0pt{#1\hss}}
\def\lta{\mathrel{\spose{\lower 3pt\hbox{$\sim$}}
    \raise 2.0pt\hbox{$<$}}}
\def\gta{\mathrel{\spose{\lower 3pt\hbox{$\sim$}}
    \raise 2.0pt\hbox{$>$}}}

\input epsf

\begin{document}

\title
	{The So-Called ``Bar'' in the Large Magellanic Cloud}

\author{HongSheng Zhao}
\affil{Sterrewacht Leiden, Postbus 9513, 2300 RA Leiden, \\
The Netherlands \\
Email: {\tt hsz@strw.leidenuniv.nl}}
\author{N.\ Wyn Evans}
\affil{Theoretical Physics, 1 Keble Road, Oxford, \\
OX1 3NP, UK \\
Email: {\tt nwe@thphys.ox.ac.uk}}

\date{Accepted ........      Received .......;      in original form .......}
\label{firstpage}

\begin{abstract}
We propose that the off-centered ``bar'' in the Large Magellanic Cloud
(LMC) is an unvirialized structure slightly misaligned with, and offset from,
the plane of the LMC disk.  The small displacement and misalignment are
consequences of recent tidal interactions with the SMC and the Galaxy.
This proposal, though radical, is consistent with the kinematics of
the LMC and near-infrared star count maps from the DENIS and 2MASS
surveys.  It does not violate any of the observational limits on the
depth and structure of the LMC -- in particular, the reported
$25^\circ-50^\circ$ inclination range of the LMC and the east-west
gradient of distance moduli of standard candles.  Contributions to LMC
microlensing come from the mutual lensing of stars in the disk and the
``bar'', as well as self-lensing of the disk and the ``bar''.  The
microlensing optical depth of such configurations from self-lensing
alone lies between $0.5 - 1.5 \times 10^{-7}$, which is comparable to
the observed microlensing signal reported by the EROS and MACHO
groups.  Possible observations are suggested to discriminate between
our misaligned, offset ``bar'' model and the conventional picture of
an off-centered, planar bar.
\end{abstract}

\keywords{Galaxy: halo - Galaxy: kinematics and dynamics - Magellanic
Clouds - galaxies: interactions - dark matter}

\section{Introduction}

By now, self-lensing is very well-established as a main cause of the
high optical depths towards the Galactic Center (Kiraga \& Paczy\'nski
1994, Zhao, Spergel \& Rich 1995, Evans 1995).  Also, there is ample
evidence that the Small Magellanic Cloud (SMC) itself provides the
lenses for the two events (MACHO 97-SMC-1 and MACHO 98-SMC-1) seen in
this direction (Alcock et al. 1997, Afonso et al. 1998, Sahu \& Sahu
1998). So, the Large Magellanic Cloud (LMC) seems something of an
anomaly, as the estimates of its self-lensing optical depth appear to
be too low to account fully for the entire microlensing optical depth
(Gould 1995). Even so, there are tantalising hints that perhaps two of
the events do indeed reside within the LMC, namely MACHO-LMC-1a and
MACHO-LMC-9. The source star for MACHO-LMC-1a lies in an
underpopulated region of the H-R diagram just below, and to the red
side of, the red clump. One possible resolution of this difficulty is
that the source star has been reddened and/or lies a few kpc behind
the LMC disk. This configuration seems more natural if self-lensing is
appreciable (Zhao, Graff \& Guhathakurta 2000).  The event MACHO-LMC-9
is a binary caustic crossing event and the projected velocity of the
lens is very low, again suggesting that it resides within the LMC
itself (Kerins \& Evans 1999).

The most striking feature of the LMC is an off-centered ``bar'' in the
optical and infrared.  Many of the computations of self-lensing in the
LMC make the implicit assumption of a co-planar thin ``bar'' and disk,
which leaves little room for star-star lensing in the LMC (e.g., Sahu
1994, Gyuk, Dalal \& Griest 2000).  Observationally, the thickness of
the LMC is not measured more accurately than $0.05$ mag using the
distance moduli of LMC standard candles.  The co-planarity of the
``bar'' and the disk is established to no better than the uncertainty
in the inclination of the LMC, which is between $25^\circ$ and
$50^\circ$ (Westerlund 1997).  This means that the ``bar'' may be
misaligned with the plane of the LMC disk by $\lta 25^\circ$ and
offset along the line of sight direction by $\lta 2$ kpc, for example.

At first sight, the introduction of misalignments and vertical offsets
may seem extreme. But, it must be remembered that the conventional
picture is not without its own problems.  In particular, it seems
contrived to require strict coplanarity and exactly zero offset in the
line of sight direction, having conceded off-centeredness on the sky
plane.  De Vaucouleurs \& Freeman (1972) named an entire class of
galaxies -- the Magellanic irregulars -- after the LMC and SMC because
of their mysterious, asymmetric bars and irregular morphology.  It is
now widely accepted that the irregular ``bar'' of the SMC is in fact
the projection of a disrupting satellite, severely stretched along the
line of sight; our polemical suggestion is that the LMC ``bar'' may
also be an illusory artifact of projection.  The LMC ``bar'' and the
SMC ``bar'' are comparable in solid angle, luminosity, age and metalicity, and reside in a common
environment in which they are continuously shaken by the tidal forces
of the Galaxy.

The idea that the structure of the LMC may be described by mass
concentrations in different planes has been suggested before (e.g.,
Johnson 1959, McGee \& Milton 1966).  There is also tantalising evidence 
for a kinematic sub-component in the LMC carbon star sample 
moving $30\kms$ relative to the LMC (Graff et al. 2000), resembling
the subcomponent seen in 21-cm channel maps of LMC (Luks \& Rolhfs 1992).
Here, we give specific models of
the LMC with the planes of the disk and the ``bar'' being misaligned,
and show that they are in good agreement with present data of the LMC
(\S 2).  Even better, we find that mild misalignment substantially
enhances the microlensing optical depth to values that are already
comparable with the observations (\S 3).  We return to the question of
origin and sketch some likely evolutionary pathways that could give
rise to the proposed misalignment (\S 4). Finally, predictions and
tests of our ideas are also discussed.

\section{Misaligned Models and Consistency Checks}

We adopt conventional models for the parameterisation of the surface
density of the disk and the ``bar''.  We require the models reproduce
the overall morphology of the LMC in projection.  Our only
modification is that we do not insist that these distributions lie in
the same plane.

There is general concensus on the location of the center of the
``bar''.  De Vaucouleurs \& Freeman (1972) take it to be
$(5^h24^m,-69.8^\circ)$, which is the center of the optical isophotes
of the bright ``bar''. There is less agreement on the center of the
LMC disk.  There are a number of population centroids within
$0.5^\circ-2^\circ$ of the optical center of the ``bar'' (see Table
3.8 of Westerlund 1997).  They are all approximately to the north of
the ``bar'', with significant scatter in the east-west directions.
For our default model, we adopt a conservatively small offset of
$1^\circ$ to the north-west in the sky plane.  The offset along the
line of sight will be dealt with later.  We set up a rectangular
coordinate system $(X,Y,Z)$ with $Z$ being the line of sight direction
through the optical center of the LMC ``bar'', $X$ being the direction
of decreasing right ascension and $Y$ being increasing declination.

The star count density of the disk in the sky plane is parameterised as
a standard exponential disk with elliptical contours
\begin{equation}\label{disksurf}
\Id(\xd, \yd) = {\NLMC (1-\fb) \over 2\pi \Ld \Wd}
\exp \left( - \sqrt{ {\xd^2 \over \Ld^2} + {\yd^2 \over \Wd^2} }
\right),
\end{equation}
while the ``bar'' is parameterised to have boxy contours and sharp edges
\begin{equation}\label{barsurf}
\Ib(\xb,\yb) = {\NLMC \fb \over 3.286 \Lb \Wb} \exp 
\left(- {\xb^4 \over \Lb^4} - {\yb^4 \over \Wb^4} \right),
\end{equation}
where the coordinates $(\xd,\yd)$ are rotated from $(X,Y)$ by the
required position angle to coincide with the apparent major and minor
axes of the disk, which have characteristic length and width
$(\Ld,\Wd)$.  Similarly, $(\xb,\yb)$ are the rotated and displaced
axes while $(\Lb,\Wb)$ are the scales of the ``bar''.  These
parameterizations are motivated by the appearance of the ``bar'' and
the disk in the optical bands and in the DENIS J, H, K bands (e.g.,
Cioni, Habing \& Israel 2000).  The model parameters are given in
Table~1.  They are designed to reproduce a nearly round outer disk
with the line of nodes at position angle $170^\circ$, and a $1\times
3$ kpc inner ``bar'' at position angle $120^\circ$.  The model is
normalised by $\NLMC$, the total number of stars of the LMC.  The
fractions of stars in the ``bar'' $\fb$ and the disk $1- \fb$ are
allowed to vary.

The value of the inclination angle of the LMC has been a long-standing
puzzle. It is roughly constrained by various tracers to lie between
$25^\circ-50^\circ$ (see Table 3.5 of Westerlund 1997); the most
recent value from the DENIS survey (Weinberg \& Nikolaev 2000) is
$\sim 42^\circ\pm 7^\circ$.  The inclination results depend on the
techniques used, whether surface brightness map fitting, deprojecting
kinematic maps, or fitting the standard candles. In all cases, the LMC
disk has been assumed to be circularly-symmetric and razor-thin.
Given the highly asymmetric distribution of stars and gas, part of the
differences in the results may well be explained by the existence of
multiple planes.

In our misaligned model, the ``bar'' and the LMC disk are treated as
two inclined slabs with their distances increasing from east to west.
One slab, be it the disk or the ``bar'', is inclined by $25^\circ$,
and the other by $50^\circ$.  The separation of the mid-planes of the
two slabs grows linearly with the projected coordinates $(X,Y)$, so
the two planes differ in the line of sight distance by an amount
\begin{equation}\label{Z0}
\Delta_{\rm bd} = |Z_0 + c_1 X + c_2 Y|,
\end{equation}
where the dimensionless constants $c_1$ and $c_2$ are determined by
the inclination of the planes, while $Z_0$ is the amount by which the
``bar'' is elevated from the LMC disk in the line of sight direction.
A more sophisticated treatment of the three-dimensional and generally
triaxial structures of the ``bar'' and the disk would introduce
additional observationally unconstrained parameters without obvious
gain in insight to the problem.  Motivated by the observation that the
distance moduli in the LMC show a predominantly east-west gradient,
here we adopt
\begin{equation}
c_1  \sim \tan(50^\circ) - \tan(25^\circ) \sim 0.7, \qquad c_2 \sim 0.
\end{equation}
We let the elevation $Z_0$ of the ``bar'' to vanish in our most
conservative model, but explore the effects of a modest level of
displacement of the ``bar'' from the disk, $Z_0 \sim \pm 1\kpc$,
allowed by the uncertainties in the distance moduli.

\begin{figure}
\epsfysize=15cm \centerline{\epsfbox{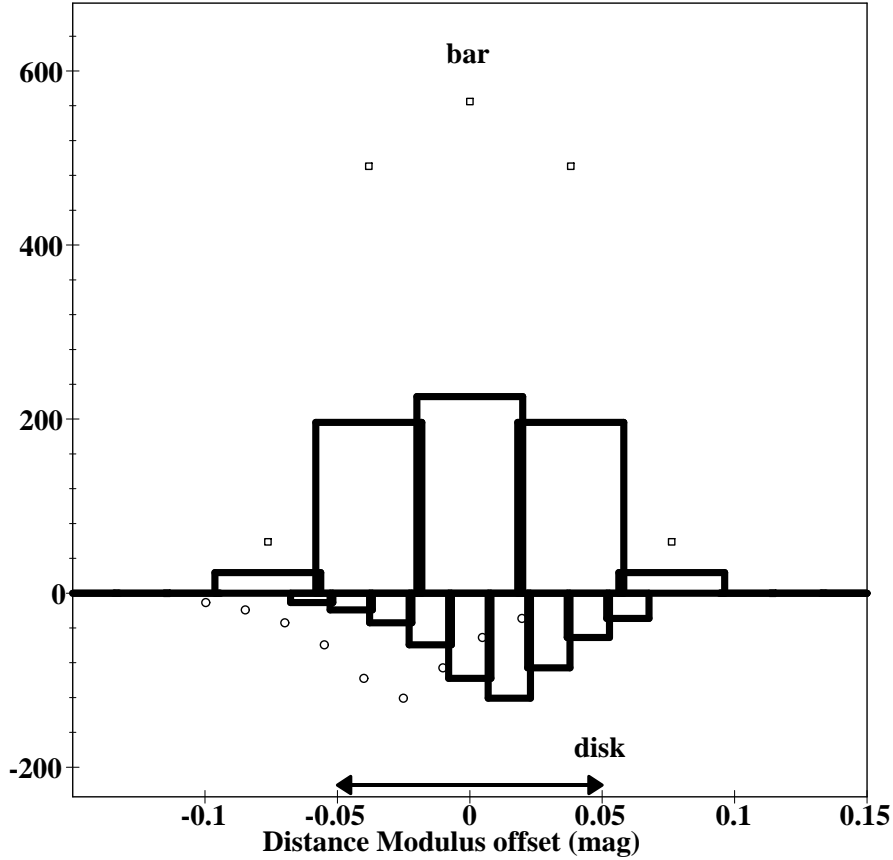}}
\caption{This shows the distance distribution of stars along the $Y=0$
line in our default models in terms of the relative distance modulus
from the LMC bar center.  Each bin is for a line of sight with a
different RA in steps of $6.4$ minutes (equivalent to about 0.8 kpc in
the X direction).  The distribution of the disk (inverted bins) is
shown separately from that of the bar (the fatter, more spaced out
bins).  The heights of the bins give the surface densities
(in$\solmasspc$) at these impact parameters. The bins are drawn for a
low-mass model with $\Mb=1\times 10^{9}\msun$, $\Md=2.5\times
10^{9}\msun$ and $Z_0=0$.  For a high-mass model with
$\Md=\Mb=2.5\times 10^{9}\msun$ and $Z_0=1\kpc$, the bins move to the
positions indicated by small circles for the disk stars and small
boxes for the bar stars.  Plotted as a double-headed arrow for
comparison is a characteristic $\pm 0.05$ mag. error bar for standard
candles. }
\label{fig:distmoduli}
\end{figure}

To check whether our model is consistent with the basic observational
data on the LMC, we compute the surface density maps and distribution
of distance moduli of the LMC stars for our default model with
parameters given in Table~1.  Figure~\ref{fig:distmoduli} shows that
the distance moduli have a spread within 0.1 magnitude for the entire
LMC, and even smaller for any given line of sight.  For comparison,
the spread in distance moduli of the RR Lyraes is $\sim 0.7$
magnitudes and so is much larger (Graham 1977, Nemec, Hesser \& Ugarte
1985, Westerlund 1997).  These authors suggest that this may be
evidence for a significant depth effect, although metallicity effects
may be another possibility.  Of course, the locations of the fields
examined by these authors are far from the bulge ($>4^\circ$) and so
they have no direct bearing on the issue of a significant depth effect
in the central region.  Our main points are that the scatter in
distance moduli produced by misaligned planes in our model is well
within the observational constraints and that there is no conclusive
evidence for a planar distribution other than theoretical
simplification.

The surface density of all stars in the LMC is calculated via
\begin{equation}\label{sigma}
\Sigma(X,Y) = {\left(\Ib + \Id\right) \MLMC \over \NLMC},
\end{equation}
where we assume a spatially constant conversion factor $\MLMC/\NLMC$
from the star count density of the disk and the ``bar''.  The constant
$\NLMC$ drops out of the calculation of $\Sigma$ eventually because of
scaling of the star count densities of the ``bar'' and the disk (see
eqns~\ref{disksurf} and~\ref{barsurf}), i.e., $\Ib/\NLMC \propto f_b
\equiv \Mb/\MLMC$ and $\Id/\NLMC \propto (1-f_b) \equiv \Md/\MLMC$,
where $\Mb$ and $\Md$ are the mass of the ``bar'' and the disk
respectively.  Figure~\ref{fig:surfdensity} shows logarithmic contours
of surface density in unbroken lines. This resembles
Figures 2 - 4 of Cioni et al. (2000), which show star counts from the
DENIS survey.

\section{Microlensing Implications of the Misaligned Disk and ``Bar''}

The microlensing map is the contour plot of the optical depth (e.g.,
Evans 1994).  In the limit that the source and the lens are at roughly
the same distance, it is calculated via
\begin{equation}\label{tau}
\tau(X,Y) \sim 10^{-7}  \times
{\Sigma(X,Y) \over 160 \solmasspc} \times {\Delta(X,Y) \over 1 \kpc}.
\end{equation}
Here, the factor
\begin{equation}\label{delta}
\Delta(X,Y) =  {\Ib^2 \hb + \Id^2 \hd + 
\Ib \Id {\rm Max} \left(\hb+\hd,\Delta_{\rm bd}\right)
\over \left(\Ib + \Id\right)^2},
\end{equation}
is the average separation between the source and the lens, where the
three terms in the numerator account for the self-lensing of the
``bar'', the self-lensing of the disk and the mutual lensing between
the ``bar'' and the disk respectively.  The depth parameters for
self-lensing of the ``bar'' and disk components are $\hd$ and $\hb$.
We set $\hb \sim 0.4\hb \sim 0.1 \kpc$ (cf. Table~1).  This appears
fairly conservative as well, given that Weinberg \& Nikolaev (2000)
detected an intrinsic spread of a few kpc in distance among their
2MASS sample of the LMC disk and bar.

The observed microlensing optical depth is $\tau_{\rm
obs}=1.1^{+0.76}_{-0.5} \times 10^{-7}$ at 95\% confidence level
(Alcock et al. 2000); the contribution from stellar lenses in the
Milky Way disk and spheroid is only $\sim 10^{-8}$.  Shown in
Figure~\ref{fig:surfdensity} in dashed lines are the contours marking
the model optical depths:
\begin{equation}
\tau = \left(1,~2,~3\right) \times 10^{-7} 
\times \left( {\MLMC \over 10^{10} \msun} \right) + 10^{-8}.
\end{equation}
The regions enclosed by the optical depth contours encompass the
locations of most of the microlensing events (the marked circles).
This is because the typical density $\Sigma$ (thin solid contours)
near the events is in the range $(160-640)\solmasspc \times \left(
{\MLMC / 10^{10} \msun} \right)$ independent of the division of the
``bar'' and the disk.  So eq.~(\ref{tau}) predicts {\it significant
star-star lensing as long as observations allow for a modest
dispersion in distance moduli} (of the order 0.05 mag, which
corresponds to 1 kpc in distances of the LMC stars).  This is a
general, robust result, insensitive to the exact division of mass
between the ``bar'' and the disk and the details of the
three-dimensional structure of the LMC.  This is verified by
calculating 100 models, which are drawn randomly with the disk-bar
offset between $\pm 1.5 \kpc$ in the $X$, $Y$ and $Z$ directions, the
inclinations of the bar and the disk between $25^\circ$ and
$50^\circ$, the thicknesses in the range $0 \le \hd=0.4\hb \le 0.1
\kpc$, and the lensable mass in the range $2.5\times 10^9 \le \MLMC
\le 5.5\times 10^9\msun$ with the bar making up between 25\% to 50\% of
it.  These models span the likely range of the three-dimensional
structure of the LMC. The average of these models (the thickest solid
line in Figure~\ref{fig:cuts}) is consistent with the observed optical
depth.  The spatial profile of the optical depth is usually
asymmetric, and changes with the bar-disk division $\Mb/\Md$, the
$(X,Y,Z)$ offset of the disk, and the inclination of the bar.  Models
with vanishing vertical offset ($Z_0=0$) generally produce an optical
depth map similar to that shown in Figure~\ref{fig:surfdensity},
except that the magnitude of the optical depth and its asymmetry are
often reduced.

Our optical depth is proportional to the total lensable mass of
$\MLMC=\Mb+\Md$ of the LMC disk and ``bar'', so that the reader can
easily adjust our optical depth values to any preferred mass.  Earlier
estimates of the dynamical mass of the LMC range between $0.6-2
\times 10^{10} \msun$ (see Table 3.4 of Westerlund 1997).  These
include a small amount of gas and a possible WIMP halo of the LMC, and
should only be taken as an upper limit to the stellar mass of the LMC.
Recent data of Kim et al. (1998) suggest a dynamical mass $\le 3.5
\times 10^{9} \msun$ inside 4 kpc radius, and a total disk mass $\sim
2.5 \times 10^{9} \msun$.  Interestingly, even for our conservative,
low-mass models (cf. Table~1), there are still enough lenses in the
LMC to account for the observed optical depth at about the $2\sigma$
level (the dashed curves in Figure~\ref{fig:cuts}).

\section{Discussion and Conclusions}

This {\it Letter} has argued that the LMC might have a radically
different structure than the conventional picture of an in-plane,
off-centered ``bar''. Our modelling starts from the premise that the
planes of the LMC ``bar'' and disk are misaligned. The centers of the
bar and disk are offset both in the sky plane and along the line of
sight. We have established first, that the combined surface density
distribution looks like the LMC as revealed by star counts in the
DENIS survey and second, that the self-lensing optical depth lies
within the interesting range ($\sim 1 \times 10^{-7}$).  Of course,
the quantitative prediction changes with the mass ratio of bar to
disk. But, in all the cases we have investigated, misalignment can
produce a significant fraction ($\gta 50 \%$) of the mean observed
optical depth.  The point is that self-lensing is insensitive to the
details of the mass ratio, but is very sensitive to the relative
separation of the LMC ``bar'' and disk stars.  Our conclusion of
significant self-lensing is a unifying one, as microlensing towards
the LMC itself then falls into line with what is already
well-established for the Galactic bulge and the SMC. In principle
there are enough lenses in the LMC to provide some (perhaps most) of
the observed optical depth.  If we think of star-star lensing as noise
on top of the signal from Macho-star lensing, then the noise is
comparable to the signal.  The significant uncertainty of the measured
inclinations of the ``bar'' and the disk does not allow us to exclude
the possibility of all lenses being from the LMC.

Now let us present two interpretations of our misaligned geometry.
This is clearly a speculative matter, and -- before we begin -- let us
emphasise that even the conventional picture of an in-plane,
off-centered ``bar'' lacks a convincing evolutionary pathway.  First,
the ``bar'' could be a tidally stretched companion of the LMC,
originating from the proto-Magellanic Cloud.  Current data and models
(e.g., Gardiner, Sawo \& Fujimoto 1994, Zhao 1998)
of the Magellanic Stream and objects along its great circle support
the hypothesis, first made by Lynden-Bell (1982), that the LMC has a
gas-rich progenitor with one or several small companions of different
star formation and chemical evolution histories.  The SMC was probably
streched and released from the grasp of the LMC by tidal interactions
only 200-500 Myrs ago, judging from its 20 kpc separation from the
LMC.  Ursa Minor and Draco must have been released at much earlier
times, if they were also companions of the proto-Magellanic Cloud.
Prior to its very recent release, the SMC must have been a tidally
stretched companion within a few kpc of the LMC, resembling the
present off-centered ``bar'' of the LMC in projection.  
This suggests the ``bar'' 
may be a cousin of the SMC, the only differences being that the
``bar'' is still within the grasp of the LMC and that the ``bar'' has
slightly higher metallicity and total luminosity than the SMC.  A
possible objection with this scenario is that it takes roughly $\sim
200 \times (10^9\msun /M) $ Myr for a lump of mass $M$ to spiral in to
the center, on account of dynamical friction provided by the halo of
the LMC (if it exists).  We only remark that the very existence of the
SMC argues empirically that somehow the multi-body dynamics makes it
possible for massive lumps to survive within the potential well of the LMC for
most of the Hubble time.  A second possibility is that
the ``bar'' may be genuine, but offset and tilted from the plane of
the disk.  This works only if both ``bar'' and disk are dynamically
young, since dynamical friction is likely to enforce co-planarity on
timescales of a rotation period (100 Myr at 1 kpc).

All these interpretations are subject to constraints from the absence
of any large offset in the radial velocity and the distance of the
``bar'' and the LMC disk.  This is not a serious problem if the
``bar'' is a real bar with its major axis tilted off the plane of the
LMC disk, and its center tied to the mid-plane of the disk.  If the
``bar'' is the projection of an unviralized companion of the LMC, then
such an object is likely to be found within a few kpc of the LMC as a
consequence of dynamical friction and orbital decay.  Such
configurations are efficient providers of star-star microlensing
because the elevation of the ``bar'' from the disk $Z_0$ is non-zero
in general, and $\tau$ is roughly proportional to $|Z_0|$
(cf. eqns.~\ref{Z0} and~\ref{tau}).  As for the velocity constraint,
any tidally stretched lump could be moving primarily in the transverse
direction across the sky in either direction along the elongated
``bar''.  As a result, the radial velocity could be close to the
systematic velocity of the LMC.  Any gradient of radial velocity
across the ``bar'' is likely small, and might mimic the velocity
pattern of the low-amplitude ($\le 30 \kms$), roughly solid-body
rotation observed in the ``bar'' region.  Given the significant
irregularities in the observed velocity field, it is far from clear
whether the conventional explanation as rotation in a nearly face-on
asymmetric disk is unique.

\begin{figure}
\epsfysize=15cm \centerline{\epsfbox{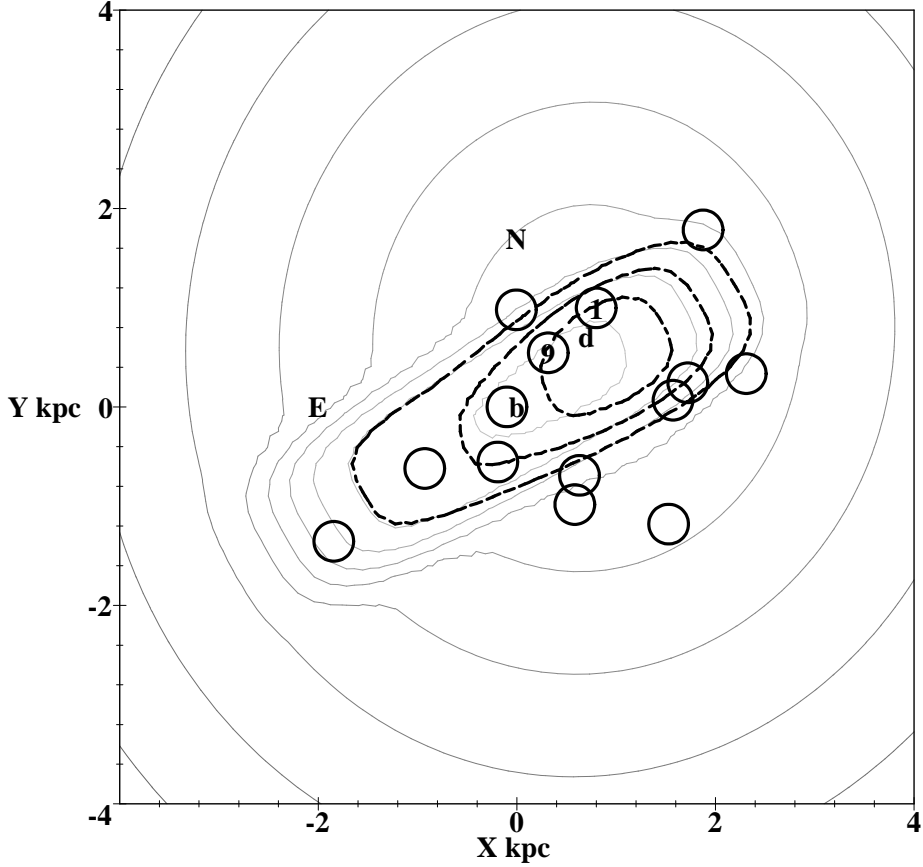}}
\caption{This shows the surface density contours of our high-mass
model (with the elevation $Z_0=1\kpc$, see Table~1) in steps of a
factor of two with the contour passing through the letter ``N'' being
$160\solmasspc \times (\MLMC / 10^{10}\msun)$.  The centers of the bar
and the disk are marked by ``b'' and ``d''.  Also shown are the
optical depth contours (dashed lines) at $\tau = (1,2,3) \times
10^{-7} \times (\MLMC / 10^{10}\msun)+10^{-8}$.  The positions of the
14 most certain events from the MACHO survey are also indicated as
circles with the two self-lensing suspects, the events MACHO-LMC-1a
and MACHO-LMC-9 labelled as ``1'' and ``9''.  These events, except for
MACHO-LMC-9, are from the sample A of Alcock et al. (2000).  Note
$1\kpc \sim 1^\circ$ at the distance of the LMC, and East is to the
left.  }
\label{fig:surfdensity}
\end{figure}

\begin{figure}
\epsfysize=15cm \centerline{\epsfbox{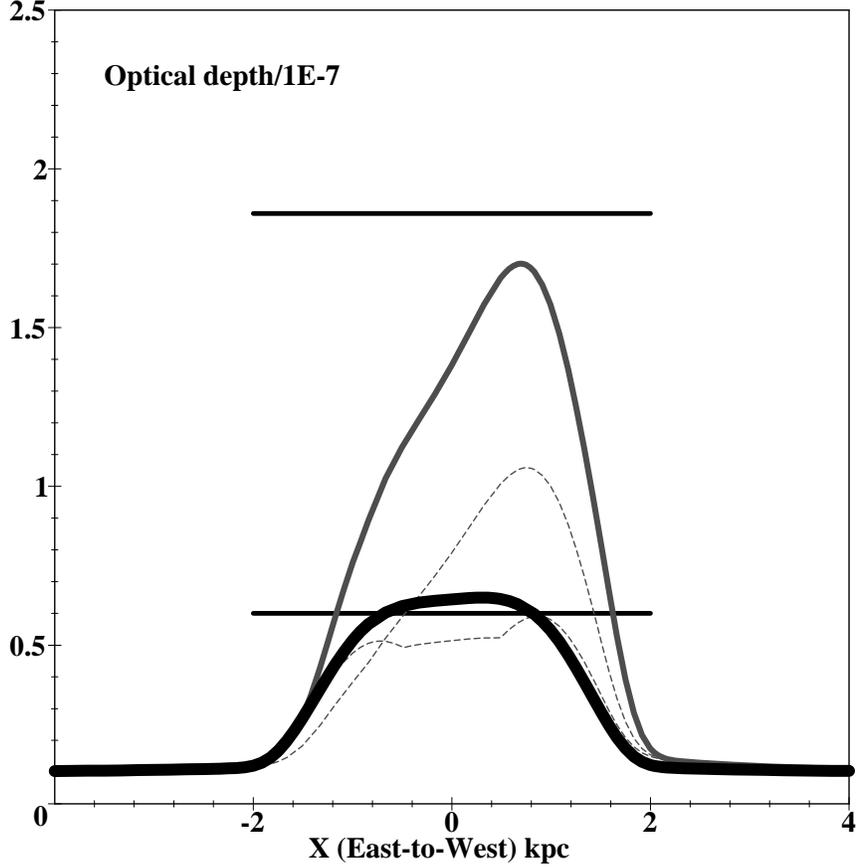}}
\caption{This compares the run of the optical depth $\tau \times 10^7$
along the decreasing RA direction (X-axis) for a number of
models. Also shown is the $95\%$ confidence range of the MACHO
observed value for events within 2 kpc of the LMC (the two horizontal
lines, Alcock et al. 2000).  The upper solid curve is for a high-mass
model with $\Mb=\Md=2.5\times 10^{9}\msun$ and the elevation
$Z_0=1\kpc$.  The two dashed curves are for low-mass models with
$\Md=2.5\times 10^{9}\msun$, $\Mb=1\times 10^{9}\msun$ and $Z_0=0$
(lower dashed curve) and $Z_0=1\kpc$ (upper dashed curve).  The
default values of the parameters are given in Table~1.  The thickest
solid line shows a robust prediction by averaging a wide range of
models allowed by observations (see text).  }
\label{fig:cuts}
\end{figure}

How can our model be falsified? There are a number of ways. First, the
reddening test proposed by Zhao (1999, 2000) is one possibility. The
source stars in our model tend to be on the back plane, thus
experiencing more reddening and extinction by dust in the fore plane
than average stars in the same field.  
MACHO-LMC-1a might be an example of such an
event.  Second, the likelihood estimator for the spatial distribution
of events developed by Evans \& Kerins (2000) is another powerful
discriminant. Note that the observed events do not respect the mirror
and point symmetries of the LMC ``bar''
(cf. Figure~\ref{fig:surfdensity}).  These spatial asymmetries may be
significant and are produced very naturally by misaligned models.
Third, analysis of the gas and stellar kinematics is a critical test for the
conventional picture of an in-plane ``bar''.  Given that the ``bar''
must dominate the gravity field in the inner parts, it is natural to
ask: why does the center of the rotation curve at
$(5^h21^m,-69.3^\circ)$ not coincide with the center of gravity of the
``bar'' in projection $(5^h24^m,-69.8^\circ)$ ?  In our picture, there
is no difficulty in explaining this at all.  If tides and interactions
are ultimately responsible for the disturbed appearance of the LMC and
for the offset between the ``bar'' and the disk in the $X$ and $Y$
directions (in the sky plane), it seems contrived to require exactly
zero offset in the $Z$ direction (line of sight).  The misaligned
models presented here serve as a general platform to explain these
offsets most naturally.  It would be interesting to test the
generality of such models with other Magellanic irregulars, such as
NGC 4027, NGC 4618 and NGC 4625.

\acknowledgments We thank James Binney and Frank Israel for a number
of helpful discussions, and Mario Mateo for a useful comment on
Magellanic irregulars.  NWE is supported by the Royal Society.  HSZ is
grateful for hospitality during visits to Oxford University, while NWE
thanks Leiden University for many kindnesses during working visits.

\begin{deluxetable}{lccccc}
\tablewidth{0pc} \tablecaption{Default parameters of our LMC models} 
\tablehead{
\colhead{Model} & \colhead{incl.} & \colhead{PA} 
& \colhead{sizes $\Delta:W:L$} 
& \colhead{center} & \colhead{mass} } \startdata 
disk & $25^\circ$ & $170^\circ$ & $0.10:1.35:1.5\kpc$
& $(5^h18^m,-69.0^\circ,50.33-Z_0\kpc)$ & $2.5\times 10^9\msun$ \nl 
``bar'' & $50^\circ$ & $120^\circ$ & $0.25:0.67:2\kpc$ &
$(5^h24^m,-69.8^\circ,50.00\kpc)$ & $[1,2.5]\times 10^9\msun$ \nl \enddata
\end{deluxetable}


{}

%
%
%

\label{lastpage}
\end{document}